\documentclass[sigconf]{acmart}
\usepackage{multirow}
\usepackage{epstopdf}
\usepackage{subfigure}
\AtBeginDocument{%
  \providecommand\BibTeX{{%
    \normalfont B\kern-0.5em{\scshape i\kern-0.25em b}\kern-0.8em\TeX}}}

\setcopyright{acmcopyright}
\copyrightyear{2021}
\acmYear{2021}
\acmDOI{10.1145/1122445.1122456}

\newcommand{\mc}[1]{\ensuremath{\mathcal{#1}}}

\begin{document}

\title{An Investigation of the Impact of COVID-19
Non-Pharmaceutical Interventions and Economic Support Policies on Foreign Exchange Markets with Explainable AI Techniques}

\author{Siyuan Liu}
\affiliation{%
  \institution{Swansea University}
  \streetaddress{Fabian Way}
  \city{Swansea}
  \country{UK}
  \postcode{SA1 8EN}
}
\email{siyuan.liu@swansea.ac.uk}

\author{Mehmet Orcun Yalcin}
\affiliation{%
  \institution{Maastricht University}
  \streetaddress{Minderbroedersberg 4-6}
  \city{LK Maastricht}
  \country{The Netherlands}
  \postcode{6211}
}
\email{m.yalcin@maastrichtuniversity.nl}

\author{Hsuan Fu}
\affiliation{%
  \institution{Universit\'{e} Laval}
  \city{Qu\'{e}bec}
  \country{Canada}
  \postcode{G1V 0A6}
}
\email{hsuan.fu@fsa.ulaval.ca}

\author{Xiuyi Fan}
\affiliation{%
  \institution{Swansea University}
  \streetaddress{Fabian Way}
  \city{Swansea}
  \country{UK}
  \postcode{SA1 8EN}
}
\email{xiuyi.fan@swansea.ac.uk}
\renewcommand{\shortauthors}{Liu et al.}

\renewcommand\footnotetextcopyrightpermission[1]{} 

\settopmatter{printacmref=false} 

\begin{abstract}
Since the onset of the the COVID-19 pandemic, many countries across the world have implemented various
non-pharmaceutical interventions (NPIs) to contain the spread of virus, as well as
economic support policies (ESPs) to save their economies. The pandemic and the associated NPIs have triggered unprecedented waves of economic shocks to the financial markets, including the foreign exchange (FX) markets. Although there are some studies exploring the impact of the NPIs and ESPs on FX markets, the relative impact of individual NPIs or ESPs has not been studied in a combined framework. In this work, we investigate the relative impact of NPIs and ESPs with Explainable AI (XAI) techniques. Experiments over exchange rate data of G10 currencies during the period from January 1, 2020 to January 13, 2021 suggest strong impacts on exchange rate markets by all measures of the strict lockdown, such as stay at home requirements, workplace closing, international travel control, and restrictions on internal movement. Yet, the impact of individual NPI and ESP can vary across different currencies. To the best of our knowledge, this is the first work that uses XAI techniques to study the relative impact of NPIs and ESPs on the FX market. The derived insights can guide governments and policymakers to make informed decisions when facing with the ongoing pandemic and a similar situation in the near future.
\end{abstract}

%



\maketitle
\pagestyle{plain}

\section{Introduction}
The COVID-19 pandemic has caused serious shocks on the global economy just within a short span of time.
Many countries across the world have imposed various non-pharmaceutical interventions (NPIs) to contain the spread of virus. For instance, China implemented an extremely stringent lockdown in Wuhan on January 23, 2020, which was lifted later on April 8, 2020; France closed its schools on March 16, 2020; South Korea banned international travelers from Hubei China on February 02, 2020; and Singapore started contact tracing on January 23, 2020. Although these NPIs have been effective in containing the spread of pandemic, they also lead to negative economic consequences at all scales. The closure of non-essential stores, restaurants, and business, and the disruptions of the global value chains cause direct revenue losses, extremely high unemployment rates, and sharp declines in personal incomes. Such influences were also reflected in the performance of financial markets. For example, on March 16, 2020, the Dow Jones Industrial Average encountered the worst percentage drop since the infamous ``Black Monday'' crash of 1987, i.e., dropped by 12.9\% in a single day. The S\&P index lost almost 12\% in the same day.

On the other hand, various economic support policies (ESPs) have been proposed and implemented to save economies, such as direct cash assistance for households, and the temporary stop of loan repayments for both individuals and businesses.
As the implementation timings of various levels of NPIs and ESPs are different for each country, one question arises: which NPI or ESP has the strongest influence on the economy? The answer to this question will shed some lights for the policymakers and also the market investors on which measures to rely on when they have to make the decisions.

In this paper, we attempt to give one answer to the question through studying the impact of NPIs and ESPs on the international economy. The exchange rates usually comove with a country's importing and exporting activities since they are direct components of and therefore highly correlated with the Gross Domestic Product (GDP)\footnote{GDP is one of the most common measures for the prosperity of each economy.}. The NPIs and ESPs during the pandemic pose widespread and long-lasting impacts on the economy via the channel of foreign exchange markets by disturbing the international trade directly in the short run, and influencing the aggregate demand indirectly in the long run.

For instance, the Australian dollar (AUD) hit \$0.59215 in exchange for 1 US dollar at the end of March 2020, which was the lowest level of the past 17 years. Note that some works have studied the impact of COVID-19 and NPIs on the FX markets. For example, \citet{aslam2020efficiency} studied the impact of COVID-19 on the efficiency of FX markets; \citet{demirgucc2020sooner} explored the early economic impact of COVID-19 NPIs; \citet{lazebnik2021spatio} assessed the economic losses caused by COVID-19 NPIs. However, the influence of individual NPIs or ESPs on the dynamic of FX markets has not been studied.

In this paper, we utilize Explainable AI (XAI) techniques to investigate the impact of NPIs and ESPs. XAI is a rising field in AI. In addition to developing AI systems that make accurate predictions, XAI systems ``explain'' their predictions to obtain insights from data. From a data science perspective, in addition to understanding ``What conclusion can be drawn from data'', XAI holds the key to answering the more important question ``Why such a conclusion is reached''. Various techniques of constructing the explanation  have been developed in XAI. One category is to compute explanations to data instances in the form of ``feature weights''. SHapley Additive expPlanations (SHAP)~\cite{lundberg2017unified} is one such method that is independent of underlying prediction models and built on a sound mathematical foundation.

In this work, we use SHAP to assess the impact of NPIs and ESPs on FX market. More specifically, we first train a Long Short-Term Memory (LSTM) prediction model to predict the exchange rates for G10 currencies using the data prior to the COVID-19 pandemic. Then we train a Random Forest (RF) model using the rate predicted by LSTM together with NPIs and ESPs to produce a refined exchange rate prediction. We then apply SHAP on the RF model to obtain the attribution of each NPI and ESP in the results of FX predictions. In such a way, we can obtain insights on which NPI or ESP measures have more contributions to FX dynamics, i.e., the appreciation or depreciation of exchange rates. To the best of our knowledge, this is the first work to study the impact of NPIs and ESPs on FX exchange markets using XAI techniques.

The remainder of the paper is organized as follows. Section 2 discusses the related literature. Section 3 presents the techniques we use for prediction and explanation. Section 4 introduces the proposed model of investigating the impact of individual NPI on FX market. Section 5 presents the experiment results. Finally, the conclusions are drawn in Section 6.

\section{Related Work}
There have been some studies conducted to explore the impact of COVID-19 NPIs and ESPs on the economic and financial systems across different perspectives. We briefly review them as follows.

\citet{demirgucc2020sooner} estimated the economic impacts of the NPIs implemented by Europe and Central Asia countries at the initial stage of the COVID-19 pandemic through tracing the economic disruptions based on the analysis of daily electricity consumption, nitrogen dioxide emission, and mobility records. Their results suggest that NPIs led to about a 10\% decline in economic activity across the region.

\citet{lazebnik2021spatio} developed an extended spatial-temporal SIR model to analyze the impact of NPIs on the pandemic spread and assessed the economic losses caused by the pandemic. Two NPIs, i.e., the duration of working and school days and various lockdown levels,  were incorporated into their model. The results based on their model and the Israeli economy suggest that 7.5 working hours alongside 4.5 school hours, or 89\% lockdown among children and 63\% lockdown among adults will achieve a balanced output, i.e., minimizing the death toll and maximising output.

\citet{mirza2020impact} evaluated the impact of COVID-19 on corporate solvency in the EU member states by introducing stress scenarios on the non-financial listed firms. A progressive increase in the probability of default, debt payback and declining coverage is reported. The results indicate that the solvency profile of all firms deteriorates. The authors further examined the possible policy interventions to sustain COVID-19. It was suggested that a hybrid support through debt and equality will be effective in the event of exacerbating business shocks to avoid a meltdown.

\citet{rizvi2020impact} assessed the impact of COVID-19 on the value of non-financial firms using a sample of 5,342 listed non-financial firms across 10 EU member states. Their findings show a significant loss in valuations across all sectors due to a possible decline in sales and increase in cost of equity. In the extreme cases, average firms in some sectors may lose up to 60\% of their intrinsic value in one year.

\citet{aslam2020efficiency} studied the efficiency of FX markets during the initial period of COVID-19 pandemic through a multifractal detrended fluctuation analysis using the exchange rate data for six currencies (AUD, CAD, CHF, EUR, GBP, and JPY). Their results demonstrate a decline in the efficiently of FX markets during the COVID-19 pandemic.

\citet{fasanya2021dynamic} examined dynamic spillovers between the COVID-19 pandemic and the global FX market. The authors analyzed the spillovers using the daily data for the period of December 31, 2019 to April 10, 2020 of six currency pairs, i.e., USD/EUR, USD/JPY, USD/CHF, USD/GBP, USD/CAD, and USD/AUD. Their findings indicate a high degree of interdependence between the COVID-19 pandemic and returns volatility.

It can be seen that there is no work conducted to analyze the impact of  NPIs and ESPs with a unified framework on the FX market, which is critical for government and policymakers to address the risks caused by current COVID-19 pandemic and possible future crisis.

\section{Background}
Before coming into the detail of the proposed model for evaluating the impact of NPIs and ESPs on FX markets, we review a few techniques used in this work.

\subsection{Random Forest}
Random Forest is a commonly-used machine learning algorithm for classification and regression problems~\cite{breiman2001random}. It starts from creating decision trees. A decision tree recursively splits data until the best partition to subset the data is found, which is typically trained through the Classification and Regression Tree (CART) algorithm ~\cite{breiman2017classification}. As decision trees are prone to problems like  bias and overfitting problems, random forest aggregate the predictions of a set of decision trees to reach a single result to reduce of risk of overfitting.

\subsection{Long Short-term Memory (LSTM)}
\begin{figure}[!h]

\centerline{
    \includegraphics[width=0.45\textwidth]{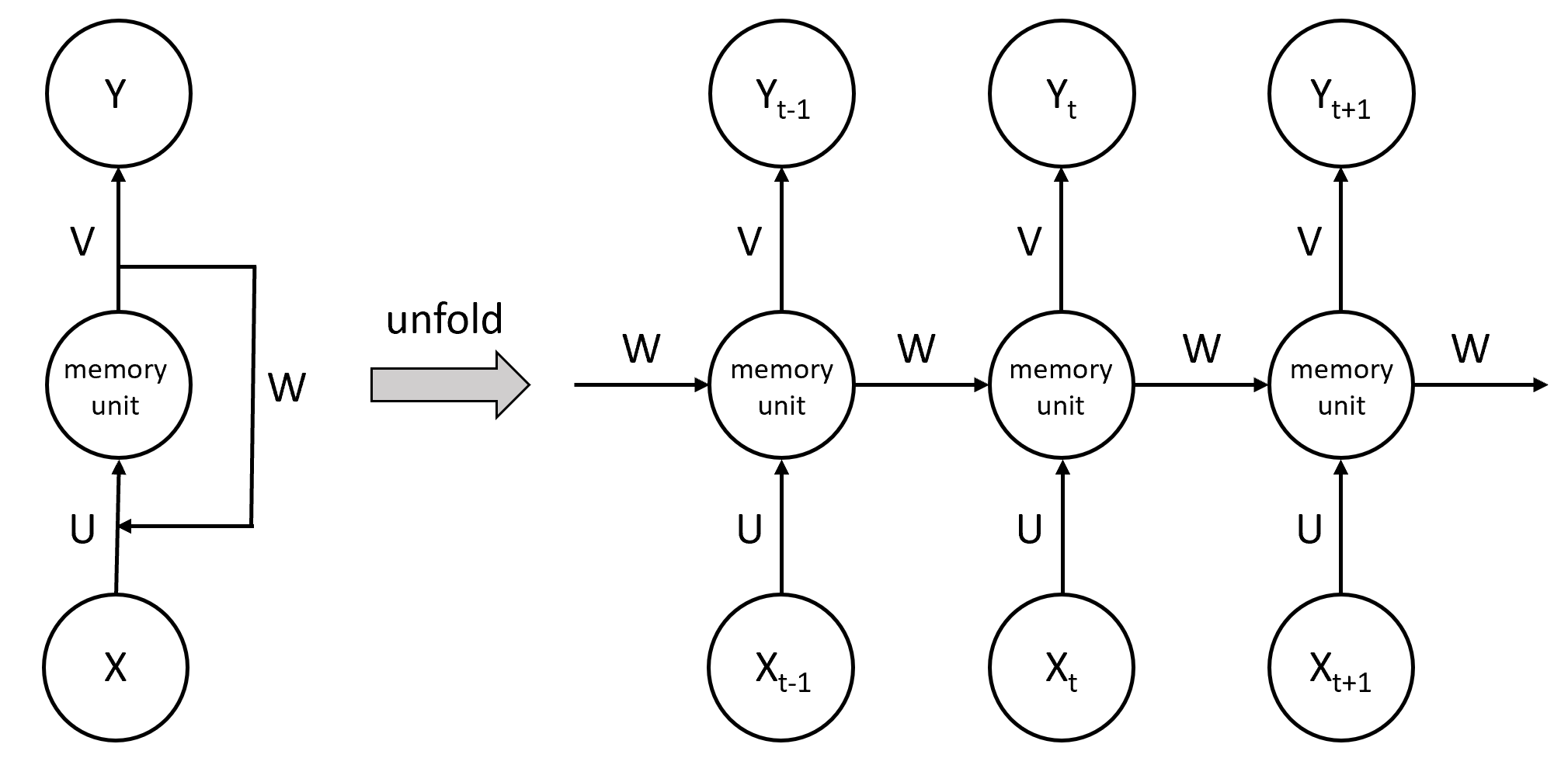}
}
    \caption{The structure of LSTM \label{fig:rnn}}
\end{figure}

Long short-term memory (LSTM) is a recurrent neural network (RNN) architecture in deep learning~\cite{hochreiter1997long}. Differing from feedforward neural networks, LSTM has connections for feedback, which makes it well-suited to classify, process and make predictions for time series data. Figure~\ref{fig:rnn} gives the overall structure of an LSTM. In the model, $X_{t}$ and $Y_{t}$ are the input and output vectors at sampling instant $t$, respectively. $U$, $W$, and $V$ are the corresponding connection weights. The structure of a memory unit depends on the variants of LSTM, such as normal LSTM and Gate Recurrent Units (GRU)~\cite{cho2014learning}.

Recently, LSTM has been widely used in FX prediction with promising results~\cite{cao2020deep}. Comparing to the traditional commonly used statistical methods, e.g., ARIMA~\cite{zhang2003time}, LSTM shows better performance when nonlinear and interconnected relationships are presented in data. Therefore, in our proposed model, we utilize a simple LSTM model to predict exchange rates. It is worth of mentioning that as the aim of the proposed model is to study the impact of NPIs and ESPs on FX market, the performance of LSTM in predicting exchange rates is not our focus.

\subsection{Shapley Explainer}
SHapley Additive exPlanations (SHAP) is a method that gives explanations to black-box machine learning predictions~\cite{lundberg2017unified}. SHAP belongs to the class of {\em feature attribution} methods. Given a prediction model $P \in \mc{P}$ where $\mc{P}$ is the set of models, let $\mathbf{y} = P(\mathbf{x})$ be the prediction
made by $P$ on the input $\mathbf{x} = \langle x_1, \ldots, x_M
\rangle \in \mathbb{R}^M$, SHAP gives an explanation $\langle \phi_1,
\ldots, \phi_M \rangle \in \mathbb{R}^M$ (for $\mathbf{y} =
P(\mathbf{x})$), where $\phi_i$ can be viewed as the contribution of
$x_i$ for this prediction.

SHAP is based on the coalitional game theory concept \emph{Shapley value}~\cite{shapley201617}. Shapley value is defined to answer the questions:``What is the fairest way for a coalition to divide its payout among the players?'' It assumes that payouts should be assigned to players in a game depending on their contribution towards total payout. In a machine learning context, feature values are ``player'', and the prediction is the ``total payout''. The Shapley value of a feature represents its contribution to the prediction and thus explains the prediction.

Specifically, let $g$ be the explanation model. For an input $x$ with $M$ features, there is a corresponding $z \in \{0,1\}^M$ such that SHAP specifies $g$ being a linear function of $z$: \[g(z) = \phi_0 + \sum_{j=1}^{M} \phi_j z_{j}\]
where $\phi_j (j > 0)$ is the Shapley value of feature $j$ and $\phi_0$ is the ``average'' prediction when none of the feature in $x$ is present. The idea is that if $z_j = 0$, the corresponding feature value is absent in $x$. Otherwise, the corresponding feature value is present in $x$.

In the context of evaluating impacts of NPIs and ESPs on FX markets, each NPI/ESP at a specific level is modelled as a feature; the predicted appreciation or depreciation of the exchange rate for a currency is the ``total payout''. By using SHAP, we get the contribution of each NPI and ESP to a prediction, implying the influence of NPIs and ESPs to the FX market, i.e., a greater contribution means a larger influence. In this work, we use the tree-based model SHAP model, TreeSHAP, for estimating Shapley values of features introduced in~\cite{lundberg2018consistent} as which is shown to be a superior method than the Kernel SHAP introduced in~\cite{lundberg2018consistent}.

\section{The Proposed LSTM-RF-SHAP Model}
Modelling NPIs, ESPs, and other factors that impact FX markets as features, we formulate the exploration of the impact of NPIs and ESPs on FX markets as evaluating the contribution of each feature to the predictions of exchange rates using SHAP.

\begin{figure}[!ht]
\centerline{
    \includegraphics[width=0.45\textwidth]{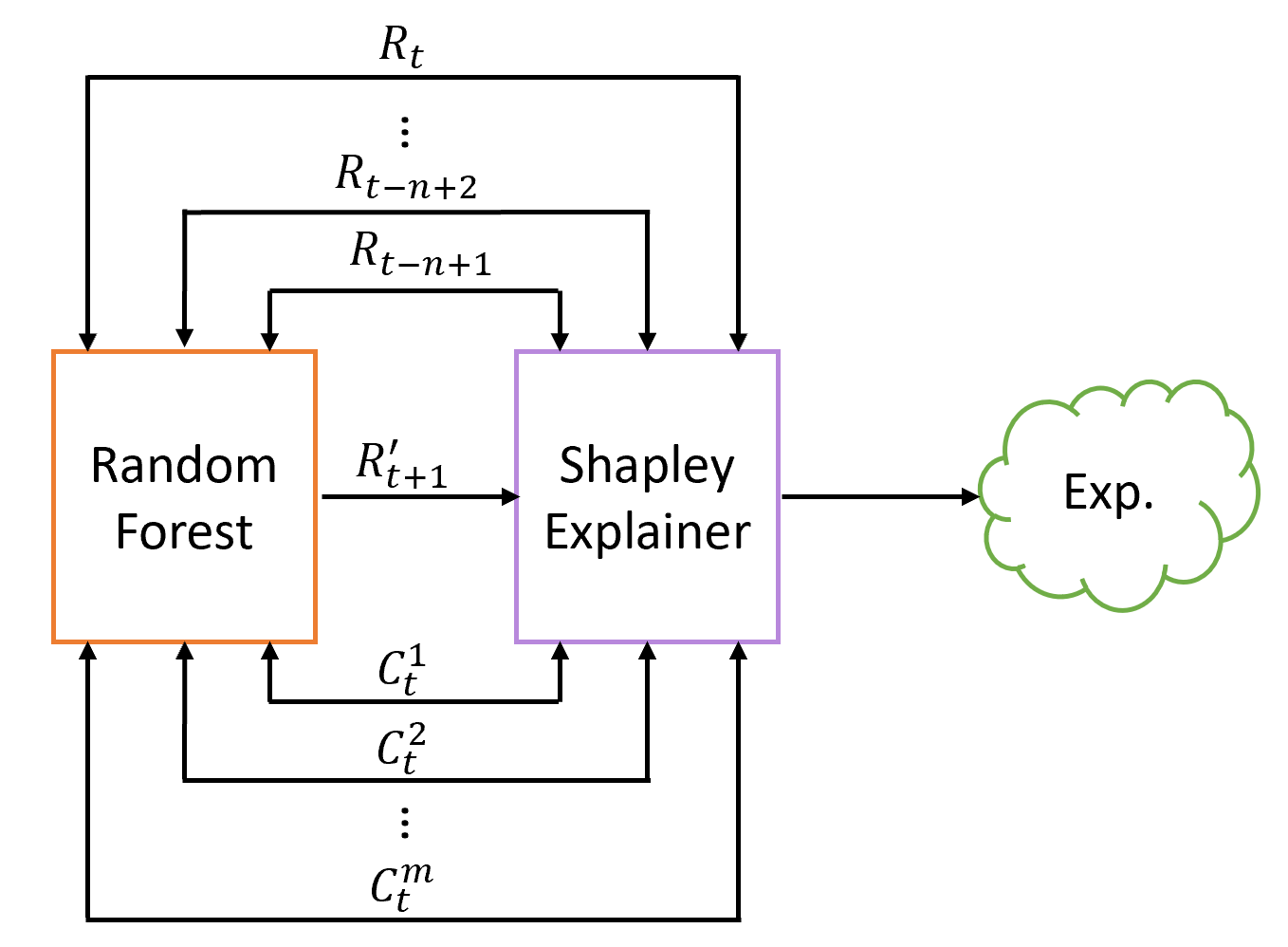}
}
    \caption{An RF-Shapley model. \label{fig:model1}}
\end{figure}

An RF-SHAP model has been initially proposed as shown in Figure~\ref{fig:model1}. In this model, suppose that we have the exchange rate for a currency\footnote{The exchange rates studied in this paper are quoted against a single currency, the US dollar.} (e.g., GBP) on day $t$, denoted as $R_t$, and $m$ features, denoted as $C^{1}_{t}$, $C^{2}_{t}$, $\ldots$, $C^{m}_{t}$. A feature can be an individual NPI or ESP, and the feature values corresponds to whether the NPI or ESP is implemented on day $t$ in the country corresponding to the currency, such as the ``2nd level lockdown'' implemented in the UK (for GBP). A feature can also be some other factors in the country that will influence the exchange rate prediction like the cumulative number of COVID-19 infection cases over last five days in the UK.

\begin{figure*}[!ht]
\centerline{
    \includegraphics[width=0.7\textwidth]{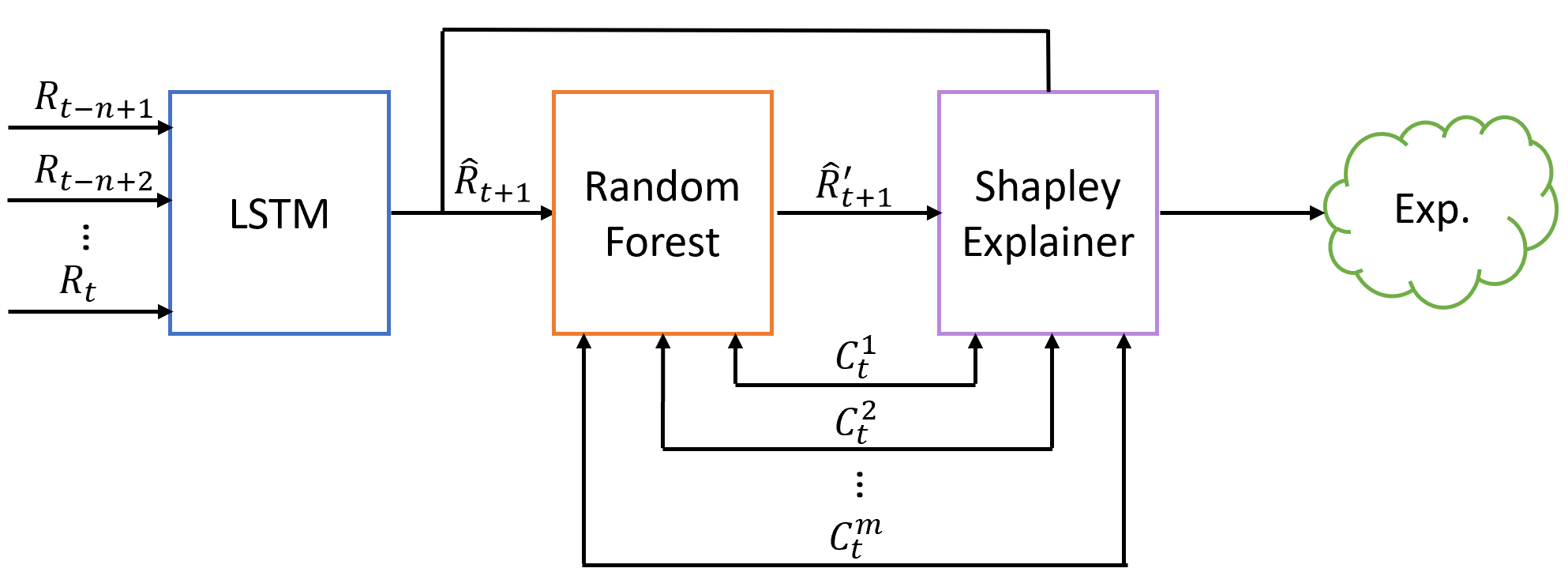}
}
    \caption{The proposed LSTM-RF-Shapley model. \label{fig:model}}
\end{figure*}
In this RF-SHAP model, we build a random forest predictor which takes inputs including exchange rates in the past $n$ days ($n$ is window size), i.e., $R_{t-n+1}$, $R_{t-n+2}$, $\ldots$, $R_{t}$, NPIs and ESPs values on day $t$, i.e., $C^{1}_{t}$, $C^{2}_{t}$,  $\ldots$, $C^{m}_{t}$,  to predict the exchange rate at day $t+1$, denoted as $R_{t+1}$. For example, suppose we are going to predict the exchange rate between GBP and USD for September 21, 2021, and the window size is five days, i.e., $n=5$. We use the exchange rates from September 16, 17, 18, 19, and 20 of the year 2021, the NPIs and ESPs values in the UK on that day, (such as that the ``Level 1 stay at home requirements'' has been enforced) on September 20, 2021, and other factors, e.g, the cumulative number of infected cases to produce the prediction. With the exchange rates being predicted from random forest, SHAP is applied to calculate the contributions of each feature, i.e., $R_{t}$, $C^{1}_{t}$,$C^{2}_{t}$, $\ldots$, $C^{m}_{t}$.

However, the contribution for each NPI and ESP obtained from this model may not reflect their real impact on the FX prediction as the actual exchange rates on $t$ are already influenced by the implemented NPIs and ESPs. Therefore, we need to disentangle the impacts of COVID-19 pandemic related factors from the non-pandemic ones. To this end, we propose a LSTM-RF-SHAP model to achieve the purpose, as shown in Figure~\ref{fig:model}.

In this LSTM-RF-SHAP model, for each currency, an LSTM model is first trained by using the historical FX data prior to 2020. Then, for day $t+1$, LSTM will take the exchange rates in the window with a size of $n$ days as input and predict the exchange rate for day $t+1$, denoted as as $\hat{R}_{t+1}$, which is independent of any information related to COVID-19 pandemic. Then this prediction together with the COVID-19 pandemic related factors will be used as an input to a random forest model, and a new predicted rate for day $t+1$, denoted as $\hat{R'}_{t+1}$, will be produced. SHAP is then applied to the random forest and we can get the contribution of each feature (i.e., $\hat{R}_{t+1}$, $C^{1}_{t}$,$C^{2}_{t}$,..., $C^{m}_{t}$) to the prediction $\hat{R'}_{t+1}$.

Taking the GBP prediction on September 21, 2021 for example, first we will train an LSTM model using the historical data before 2020, such as the exchange rates in 2019. With the LSTM model trained and window size being set at $n=$ 5, we will pass to LSTM the exchange rates of September 16, 17, 18, 19 and 20 of the year 2021 as input and will get a predicted exchange rate, i.e., $\hat{R}_{09212021}$. Then we will pass $\hat{R}_{09212021}$ together with the COVID-19 pandemic related factors on September 20, 2021, such as the NPIs and ESPs values in the UK on September 20, 2021, the cumulative number of infected cases in the past 5 days, etc. to the random forest to get the prediction $\hat{R'}_{09212021}$. SHAP is then applied to calculate the contributions of each feature, including the output from LSTM, e.g., $\hat{R}_{09212021}$, NPIs, ESPs, and other COVID-19 pandemic related factors.

\section{Experimental Settings and Results}
\subsection{Data Preparation}
We choose the G10 currencies\footnote{https://en.wikipedia.org/wiki/G10\_currencies} to study the impact of the NPIs on the FX markets as these currencies account
for over 95\% of trading volume in the worldwide FX markets~\cite{salisu2018modelling}. In particular, we collected FX data at daily frequency\footnote{The data are collected from the \textit{Datastream} database.}  for the nine currency pairs, i.e., the Australian Dollar (AUD), the Canadian Dollar (CAD), the Swiss Franc (CHF), the Euro (EUR), the British Pound (GBP), the Japanese Yen (JPY), the Norwegian Krone (NOK), the New Zealand Dollar (NZD), and the Swedish Krona (SEK), all against the US Dollar (USD) in the period of January 01, 2019 to January 13, 2021. The exchange rates from January 01, 2019 to December 31, 2019 are used to train an LSTM predictor for each currency pair. The window size to train the LSTM is set at 5. It is worth of pointing out that we will pass the predicted returns to random forest in order to improve the prediction accuracy. In particular, for a day $t$ in the period of January 1, 2020 to January 13, 2021, we will first use the exchange rate for each currency from  day $t-5$ to $t-1$ to predict the exchange rate for day $t$. Then, we compute the predicted return on day $t$ by taking the logarithm difference of exchange rates between day $t$ and day $t-1$. The exchange rates may have very different scales of market prices, e.g., 1 US dollar on January 1, 2020 can be exchanged for 108.0961 JPY versus 0.8891 EUR. The logarithm return is more comparable than the simple return across currencies as it addresses this scale issue.

Then the NPIs and ESPs values in a country and other COVID-19 pandemic related factors\footnote{The data are collected from https://github.com/OxCGRT/covid-policy-tracker}  alongside with the predicted returns from LSTM during the period of January 01, 2020 to January 13, 2021 are used to train a random forest\footnote{There are in total nine LSTM and one random forest trained.}. More specifically, the NPIs, EPSs and other COVID-19 pandemic related factors we consider are as follows.
\begin{enumerate}
\item Economic support policies: governments in different countries or regions have taken discretionary actions to sustain employment rate and solvency.
\begin{enumerate}
  \item \textbf{$E_{1}$: Income Support} -- Government provides direct cash support to people who cannot work due to COVID-19 pandemic. For example, in March 2020, the UK government implemented a policy to pay 80\% of a furloughed employee's wages (subject to a cap of GBP2,500 per month).
  \item \textbf{$E_{2}$: Level 1 Debt/Contract relief} -- Government freezes financial obligations for households (e.g., stopping loan repayments, preventing services like water from stopping, or banning evictions). Level 1 will be specific to one category of debt or contract. For example, in March 2020, the CARES Act was signed into law in United States, which implemented targeted debt relief based on Federal jurisdiction (e.g. mortgage relief).
  \item \textbf{$E_{3}$: Level 2 Debt/Contract relief} -- Comparing to Level 1 Debt/Contract relief, level 2 targets at multiple categories of debt or contract. For example, the Australian Government implemented  a series of changes to bankruptcy law in March 2020, which includes an increase in the debt threshold, an increase to the timeframe to respond to a bankruptcy notice, and an increase to the temporary debt protection period.
\end{enumerate}
\item The NPIs to contain the spread of COVID-19 pandemic
\begin{enumerate}
  \item \textbf{$N_1$: Level 1 Stay at home requirements} -- Many countries and regions implemented various levels of ``stay at home requirement''. At Level 1, it is not compulsory for residents to stay in their residences although it is recommended.
  \item \textbf{$N_2$: Level 2 Stay at home requirements} -- At this level, residents are not allowed to leave their residences without exceptions (e.g., daily exercises, grocery shopping, and essential trips).
  \item \textbf{$N_3$: Level 1 Workplace closing} --  It is recommended to close workplaces or to work from home although it is not required.
  \item \textbf{$N_4$: Level 2 Workplace closing} -- It is required to close workplaces or to work from home when possible.
  \item \textbf{$N_5$: Level 1 International travel controls} -- Since the onset of COVID19 pandemic, many countries and regions have implemented restrictions over international and internal movements. For level 1 international controls, screening (e..g., temperature taking) will be taken upon arrivals.
  \item \textbf{$N_6$: Level 2 International travel controls} -- At this level, a period (e..g, 14 days) of quarantine in designated places is required for the travellers from some countries or regions.
  \item \textbf{$N_7$: Level 3 International travel controls} -- At this level, no travellers from some countries or regions are allowed to arrive.
  \item \textbf{$N_8$: Restrictions on internal movement} -- Besides international travel controls, domestic travelling between regions is not recommended.
\end{enumerate}
  \item \textbf{$C_1$}: For other COVID-19 pandemic related factors except the ESPs and NPIs, we also take \textbf{cumulative cases in last five days} into consideration.
\end{enumerate}

\subsection{Evaluation on Prediction Accuracy}
We use directional prediction accuracy ($DA$), Mean Absolute Error ($MAE$), and Root Mean Square Error ($RMSE$) to evaluate the accuracy of the proposed LSTM-RF-SHAP model in FX predictions. Although the focus of the work is to study the impact of NPIs and ESPs on FX markets, the model performance in predicting FX price is important as well because we need to obtain insights from correct prediction instances by using SHAP, i.e., if the prediction for an instance is incorrect (e.g., the actual result is an FX appreciation but the prediction indicates rather a depreciation), we will not be able to have a meaningful analysis for the obtained explanations on such an instance.

In more detail, $DA$ is a direction measure of the FX prediction accuracy, ranges from 0 to 1, with a higher value indicating a better prediction accuracy.
\begin{equation}
DA = \frac{1}{N}\sum_{t=1}^{N}d(t)\times100\%,
\end{equation}
where
\begin{equation}
d(t)=\left\{
\begin{aligned}
&1\qquad if [y(t+1)-y(t)][\hat{y}(t+1)-y(t)]\geq0;\\
&0\qquad otherwise.
\end{aligned}
\right.
\end{equation}
where $\hat{y}(t)$ and $y(t)$ denote the predicted and the actual FX prices at day $t$, respectively, and $N$ is the number of the prediction instances (i.e., the number of working days from January 01, 2020 to January 13, 2021).

MAE is the average of the differences between the actual and the predicted FX prices where a smaller value implies a higher prediction accuracy.
\begin{equation}
MAE=\frac{1}{N}\sum_{t=1}^{N}|y_{t}-\hat{y}_{t}|,
\end{equation}
where $\hat{y}(t)$, $y(t)$, and $N$ are the same as in Equation (2).

RMSE is the standard deviation of the differences between the actual and the predicted FX prices where a smaller value denotes a better prediction performance.
\begin{equation}
RMSE=\sqrt{\frac{1}{N}\sum_{t=1}^{N}(y_{t}-\hat{y}_{t})^{2}},
\end{equation}
where $\hat{y}(t)$, $y(t)$, and N are the same as in Equation (2).

We compare the accuracy of the proposed LSTM-RF-SHAP model with Autoregressive Integrated Moving Average model (ARIMA)~\cite{zhang2003time} in the accuracy of FX predictions. ARIMA analyzes the time-series correlation and builds a prediction model from a statistical approach. We first train an ARIMA model using the exchange rates in the period of January 01, 2019 to December 31, 2019, and predict the exchange rates for a day in January 01, 2020 to January 13, 2021 using ARIMA-RF-SHAP by following the similar procedure as the proposed LSTM-RF-SHAP model.

\begin{figure*}[!ht]
\centerline{
    \includegraphics[trim=35mm 30mm 35mm 50mm,width=0.85\textwidth]{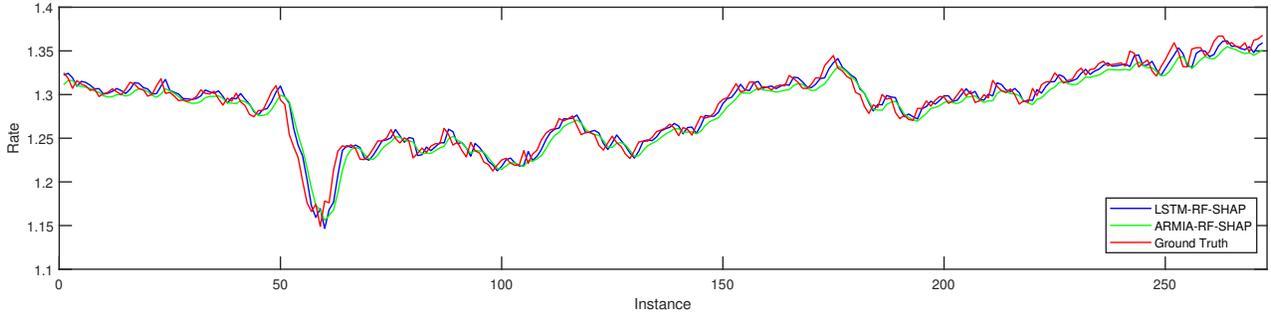}
}
    \caption{The predicted GBP/USD rates for workdays in the period of January 1, 2020 to January 13, 2021. \label{fig:accuracy}}
\end{figure*}

As an example, Figure~\ref{fig:accuracy} shows the GBP/USD exchange rate by using LSTM-RF-SHAP and ARIMA-RF-SHAP for the period of January 1, 2020 to January 13, 2021. The x-axis represents prediction instances. As there are no exchange rates available in weekends, there are 272 instances in total in this period. The y-axis represents predicted exchange rates for each instance. The blue line and green line are the predictions produced by LSTM-RF-SHAP and ARIMA-RF-SHAP for an instance, respectively. The red line shows the actual rate for the instance. It can be seen that the blue line is closer to the red line compared to the green line, suggesting that a more accurate GBP/USD rate prediction can be achieved by using LSTM-RF-SHAP.

\begin{figure*}[!ht]
\centerline{
\includegraphics[trim=20mm 0mm 20mm 0mm,width=0.85\textwidth]{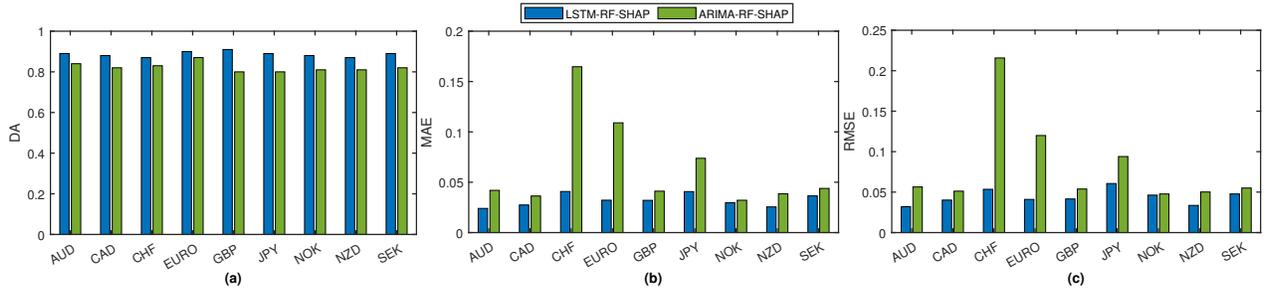}
}
    \caption{The $DA$, $MAE$, and $RMSE$ results for each currency for workdays in the period of January 1, 2020 to January 13, 2021.}
    \label{fig:accuracy2}
\end{figure*}
We further compare LSTM-RF-SHAP with ARIMA-RF-SHAP in terms of $DA$, $MAE$, and $RMSE$ for the period of January 1, 2020 to January 13, 2021. The results are shown in Figure~\ref{fig:accuracy2}. Figure~\ref{fig:accuracy2}(a), (b) and (c) show the $DA$, $MAE$, and $RMSE$ results, respectively. The blue bar represents LSTM-RF-SHAP and green bar represents ARIMA-RF-SHAP. It can be seen that the $DA$ of LSTM-RF-SHAP is consistently higher than that of AMIRM-RF-SHAP for each currency, and the errors (i.e., MAE and RMSE) are lower, suggesting that the proposed  LSTM-RF-SHAP model can achieve a more accurate FX predictions comparing to ARIMA-RF-SHAP.


\subsection{Explanations}
As we introduced in Section 4 and Section 5.1, there are 13 features in total used as the input to the random forest predictor. The 13 features are the predicted return from LSTM, the ESPs $E_{1}$, $E_{2}$, and $E_{3}$, the NPIs $N_{1}$, $\ldots$, $N_{8}$, and the number of cumulative COVID-19 cases in last 5 days $C_{1}$. SHAP is used to evaluate  contributions of each feature. To simplify the explanation task, we classify the RF prediction results as a binary exchange rate direction prediction (exchange rate appreciates or depreciates). For a day $t$ in the period of January 01, 2020 to January 13, 2021, in addition to obtaining the prediction of exchange rate appreciation or depreciation from the RF, SHAP  outputs a vector with a length of 13. The value of each element in the vector is in the range of $[-1,1]$.
If a feature has a high SHAP value, it is understood as such feature has a great impact to the FX market.

\begin{figure*}[!ht]
\centerline{
    \includegraphics[trim=10mm 0mm 10mm 0mm,width=0.95\textwidth]{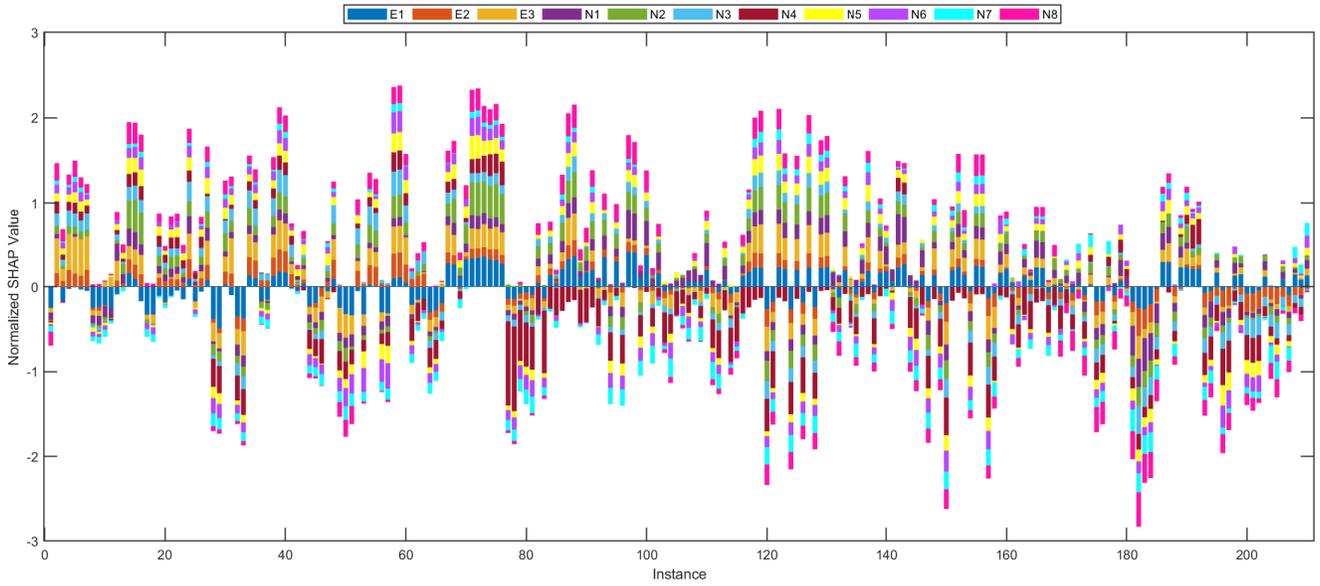}
}
    \caption{The contributions of ESPs and NPIs to correct GBP/USD prediction instances in the period of January 1, 2020 to January 13, 2021. \label{fig:uk_shap}}
\end{figure*}

As an example, Figure~\ref{fig:uk_shap} shows contributions of ESPs ($E_1$-$E_3$) and NPIs ($N_1$-$N_8$) to the predicted exchange rate directions for GBP/USD. In this figure, x-axis represents the instances; and y-axis is the normalised SHAP values. The sign of the y value on each instance is determined by the exchange rate direction of that instance. In other words, if the prediction is positive (the rate appreciates), then the y values are shown positively; otherwise, the y values are shown negatively. Only instances with correct predictions are shown in this figure; and there are 210 such instances.
The length of each color bar represents the amount of contribution made by the corresponding feature.

\begin{figure*}[t!]
\centerline{
    \includegraphics[trim=10mm 0mm 10mm 0mm,width=0.95\textwidth]{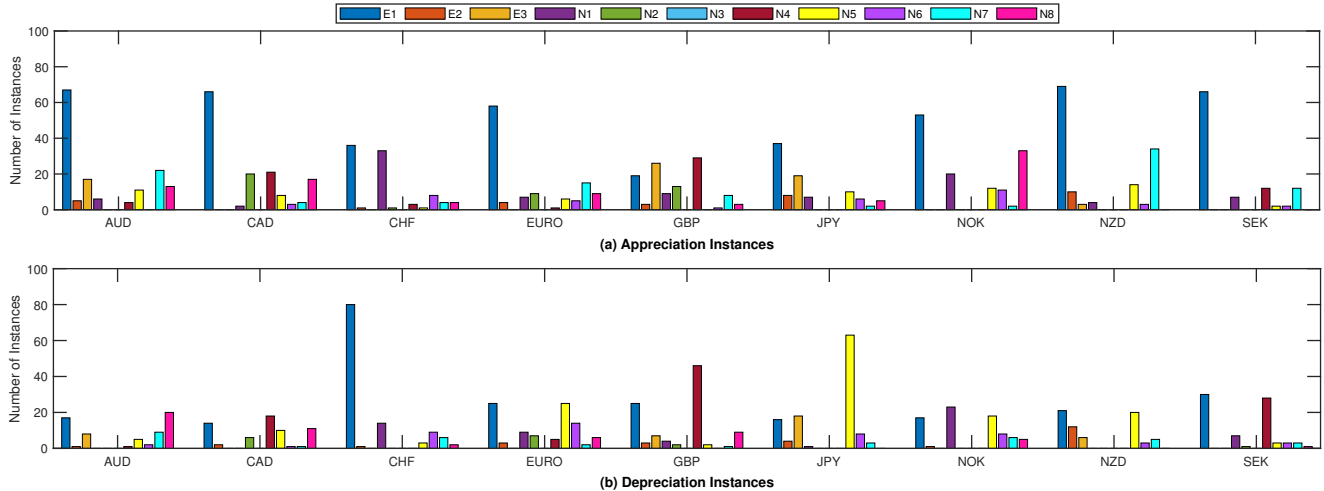}
}    \caption{The number of instances to which a particular ESP or NPI has the largest contribution in the period of January 1, 2020 to January 13, 2021: (a) Appreciation Instances; (b) Depreciation Instances. \label{fig:all_cur}}
\end{figure*}

From Figure~\ref{fig:uk_shap}, we can see that for GBP/USD, \emph{$E_{1}$ (Income support)}, \emph{$E_{3}$ (Level 2 Debt/Contract relief)} and \emph{$N_{2}$ (Level 2 Stay at home requirements)}, have the largest contributions for appreciation instances more frequently than other features. \emph{$N_{4}$ (Level 2 Workplaces closing)} has the largest contribution for depreciation instances. Note that as each feature can take different values, which include the values representing the non-existence of the control measure, so features that are prominent only indicate the ESPs and NPIs they represent are influential to the market; they do not suggest implementing such ESPs or NPIs would appreciate or depreciate the rate, as values of the features may indicate the ESPs or NPIs being not implemented.

Figure~\ref{fig:all_cur} presents the number of instances to which a particular ESP or an NPI has the largest contribution. For each currency, we separate the appreciation and depreciation instances.
Figure~\ref{fig:all_cur} (a) and (b) show the contribution for appreciation and depreciation instances, respectively. Each bar corresponds to an ESP or an NPI. The height of a bar represents the number of the instances in which an ESP or an NPI has the largest contribution.

There are some points observed when we consider the features that have greatest contribution to more than 20 instances. Firstly, from figure~\ref{fig:all_cur}(a) we can see that for most currencies except GBP, the number of appreciation instances with $E_{1}$ having the largest contribution is obviously greater than other features. This shows that {\em income support} being the most influential factor throughout all currencies. $E_{3}$ ({\em Level 2 Debt/Contract relief}), $N_{1}$ ({\em Level 1 Stay at home requirements}), $N_{8}$ ({\em Level 3 International travel controls}), and $N_{9}$ ({\em Restrictions on internal movement}) also present large contributions to some currencies.  Secondly, from figure~\ref{fig:all_cur}(b), we can see that various features present the largest contribution to the depreciation instances, such as $E_{1}$, $N_{4}$, and $N_{6}$ ({\em Level 2 International travel controls}). As indicated in the results shown in Figure~\ref{fig:uk_shap}, large values represent the presence or absence of an ESP or NPI being important to exchange rates.

\section{Conclusion}
COVID-19 pandemic and the associated non-pharmaceutical
interventions (NPIs) across the world have triggered a large size of economic shocks. Without exceptions, the global foreign exchange (FX) market is disrupted. Economic support policies (ESPs) were also implemented to boost economy. Although there has been some studies exploring the influence of the ongoing pandemic, NPIs and ESPs on FX market, the question regarding the impacts of individual ESP or NPI on the FX markets remains unanswered. In this paper, we provide one answer by using an XAI techniques, featuring attribution algorithm SHAP. To our best knowledge, this is the first work that uses XAI techniques to study the impact of individual NPI or ESP during the COVID-19 pandemic and its association with the FX markets.

In particular, we use daily exchange rate data for G10 currencies prior to the onset of COVID-19 pandemic to train LSTM models. Then for each day in the period January 01, 2020 to January 13, 2021, we first use trained LSTMs to generate predictions for exchange rates, which are subsequently fed into a random forest (RF) model alongside some COVID-19 related policy responses, such as ESPs, NPIs, as well as the number of cumulative COVID-19 cases in past days, to obtain predictions on exchange rate, either appreciation or depreciation. Later, SHAP is applied over the RF model to produce the explanations. Experimental results suggest that the ESPs, such as income support and debt/contract relief, and strict NPIs like stay at home requirements, workplace closing, international travel controls and restrictions on internal movement are associated with the appreciation and depreciation of the exchange rates. Their influences are heterogeneous across currencies.

In the future, there are a few directions we would like to explore further. Firstly, we will improve current LSTM model by incorporating macro-financial factors such as inflation rate, money supply index, consumer index, and industrial production index to achieve a more accurate prediction on the exchange rate. Achieving accurate predictions is essential for an XAI technique to produce meaningful explanations. Secondly, as we have only considered predicting and explaining exchange rates in this work, we would like to investigate other aspects of FX markets, such as efficiency, dynamic spillovers, and volatility transmissions. Lastly, we would like to apply other XAI techniques, like~\cite{lime2016,aas2019explaining}, to generate explanations to gain more comprehensive explanations.

\begin{acks}
This work is funded by the Quebec-Wales Collaboration 2020 project: {\em Understanding Impact of COVID-19 on International Economy via Currency Markets using Explainable AI.}
\end{acks}

\bibliographystyle{ACM-Reference-Format}
\bibliography{acmart}


\end{document}